\documentclass[twoside]{article}
\usepackage{fleqn,espcrc2}

\title{How  to experimentally measure the number 5 of the $SO(5)$ theory?}
\author{Jiang-Ping Hu and Shou-Cheng Zhang}

\address{\it Department of Physics, McCullough Building, Stanford University, Stanford  CA~~94305-4045}

\begin{document}

\begin{abstract}
According to Wilson's theory of critical phenomena, critical
exponents are universal functions of $d$, the dimension of
space, and $n$, the dimension of the symmetry group.
$SO(5)$ theory of antiferromagnetism and superconductivity predicts
a bicritical point where $T_N$ and $T_c$ intersect. By
measuring critical exponents close to the
bicritical point, and knowing that $d=3$, one can experimentally
measure the number $5$ of the $SO(5)$ theory.
\end{abstract}

\maketitle

%\vskip2pc

In a system of many strongly interacting degrees of freedom, it is
generally hard to make precise quantitative predictions which can
be tested experimentally. Theories of high $T_c$
superconductivity generally have to resort to uncontrolled approximations,
as a result, it is not possible for experiments
to uniquely test the fundamental physical validity of the theory.
However, at special values of physical parameters, the basic
degrees of freedom may compete so strongly that a new critical point
is reached. At this critical point, low energy properties depend
only on universal quantities such as the number of space dimension
$d$ and the dimension of the symmetry group $n$, and are
independent of the microscopic details of the constituent
materials\cite{eexpansion,review}.
Close to such critical points, a new kind of simplicity and
predictibility becomes possible, and the theoretical foundation can be
tested unambiguously by experiments.

$SO(5)$ theory predicts a new bicritical point in the two dimensional
phase diagram of temperature versus doping where the antiferromagnetic
(AF) transition temperature $T_N$ intersects the superconducting
(SC) transition temperature $T_c$. At this critical
point, the three component AF order parameter is
unified with the two component SC order
parameter to form a five component superspin order
parameter\cite{science,projected}.
At this point, the dimension of the symmetry group $n$ is enhanced to
$5$, and as a result, the critical properties at this point are
uniquely different from that of a lower symmetry world. By
experimentally tuning into such a bicritical point, and by precisely
measuring the critical exponents at this point, experiments can
therefore determine the dimension of the symmetry group and
test the fundamental validity of the $SO(5)$ theory.

The purpose of this paper is to summarize the various exponents
predicted by the $SO(5)$ theory, and to encourage experiments to
measure these exponents. Due to chemical complications, the bicritical
point has not yet been clearly identified experimentally in the high
$T_c$ cuprates. However, such a point does exist in two
dimensional organic superconductors which share many physical
properties with the high $T_c$ cuprates. Here we also review a
insightful theoretical analysis by Murakami and Nagaosa\cite{organic},
who showed that the
the NMR experiments near such a bicritical point measure the
dimension of the symmetry group $n$ to be very close to $5$.

%\section{model}
{\bf The model:} We start with  a generic Ginzburg-Laudau form of the
$SO(5)$ model,
\begin{eqnarray}
H=\frac{1}{2}\int d^{d}\mbox{\boldmath$r$}[ r_{c}|\vec{n}|^{2}+
|\vec{\nabla}\vec{n}|^{2}+r_{s}|\vec{m}|^{2}+|\vec{\nabla}\vec{m}|^{2} +
\nonumber\\
2\delta_c|\vec{n}|^{4}+
4W|\vec{n}|^{2}|\vec{m}|^{2}
+2\delta_s|\vec{m}|^{4}].
\label{model}
\end{eqnarray}
where $\vec{n}$ and $\vec{m}$ are the order parameters of the SC and
the AF respectively. In this paper, we will fix the dimension to $d= 3$
and the expansion parameter $\epsilon = 4-d =1$.
The mean field  phase diagram and RG flows of above effective
Hamiltonian have been derived in
Ref.\cite{bitetra1,bitetra2,organic}.
Defining $F =\delta_c\delta_s-W^2$, we
summarize their results in following:
{\bf (i)} When $F>0$, the RG
flow converges to the biconical fixed point $(\delta_c, W, \delta_s) =
2\pi^2(0.0905,0.0847,0.0536)$ which correponds to {\it tetracritical}
phenomena in mean field phase diagram. In this case, the AF and SC
orders can coexist in the low temperature phase;
{\bf (ii)} When $F=0$,  the RG flow converges to a
Heisenberg fixed point $\delta_c = \delta_s = W =
\frac{2\pi^2}{13}$ which corresponds to {\it bicritical} behavior.
This fixed point has an exact $SO(5)$ rotational symmetry\cite{science}.
In this case, there is a direct first order transition between AF and SC.
{\bf (iii)} When $F<0$, the RG flow goes to unstable region
$(\delta_c  \delta_s)<0)$. The first order transition line between SC
and AF branches at a triple critical point and extends until two
branches ends at tricritical points. All of above results were
discussed and summarized in  the  schematic diagrams by the authors
of Ref.\cite{organic}. In the case of $n=5$, the bicritical and the
tetracritical points are very close in parameter space. Starting from
a generic point in the parameter space, there is a rapid RG flow towards
the bicritical point, followed by a slow flow from the bicritical to
the tetracritical point. Therefore, there is a large regime of parameters
where the bicritical behavior dominates, and it is possible to observe the
$SO(5)$ symmetry. Recent Monte Carlo simulations of the classical $SO(5)$
spin models\cite{hu} are consistent with this interpretation of the
bicritical point.
%\section{Static Exponents}
%\subsection{Theory}

{\bf Static Exponents:}
We first discuss the static critical phenomena.
At the bicritical point $r_s=r_c=0$, thermodynamic quantities obey scaling
relations. However, $g=r_c-r_s$ is a relevant parameter, and the scaling
theory of a bicritical point requires a crossover critical exponent $\phi$
(Ref.\cite{spinflop}). The scaling postulate for the singular part of
the free energy takes the form ($t= |T-T_c(g=0)|$),
\begin{eqnarray}
F(t,g,\vec{n},\vec{m}) =
t^{2-\alpha}f(\frac{g}{t^{\phi}},\frac{\vec{n}}{t^{\beta}}
,\frac{\vec{m}}{t^{\beta}}).
\end{eqnarray}
The exponent $\beta$ and $\alpha$ take the  same value as  the
isotropic vector model.

The critical exponents $\alpha$, $\beta$, $\gamma$, $\delta$,
$\nu$ and $\eta$ satisfy the usual scaling relations:
%\begin{eqnarray}
%C(T) &=&at^{-\alpha}, \nonumber \\
%M(T) &=& b t^{\beta},~~B=0; ~~~~ M(B)= b'B^{\frac{1}{\delta}} ~~T=T_C ,\nonumber \\
% \chi(T)&=&ct^{-\gamma},~~  \xi = \xi_0 t^{-\nu}, \nonumber \\
% G(x) &=&dx^{-(d-2+\eta)}, ~~ G(q) = d'q^{-(2-\eta)}
%\end{eqnarray}
%Exponent relation:
\begin{eqnarray}
\alpha&=&2 -d\nu, ~~ \beta= \frac{1}{2}(d-2+\eta), \nonumber \\
\gamma&=&\nu(2-\eta),~~ \delta = \frac{d+2-\eta}{d-2+\eta}
\end{eqnarray}
Within second order $\epsilon$ expansion, they are given by
(see Ref.\cite{eexpansion} P.133 for
$\alpha$,$\beta$,$\eta$
$\nu$ and $\delta$ Ref.\cite{review} P.611 for $\phi$):
\begin{eqnarray}
\alpha &=& -\frac{(n-4)}{2(n+8)}\epsilon
  -\frac{(n+2)^2(n+28)}{4(n+8)^3}\epsilon^2 \nonumber\\
\beta &=&\frac{1}{2} -\frac{3}{2(n+8)}\epsilon
  +\frac{(n+2)(2n+1)}{2(n+8)^3}\epsilon^2 \nonumber\\
\gamma &=&1 +\frac{(n+2)}{2(n+8)}\epsilon
  +\frac{(n+2)(n^2+22n+52)}{4(n+8)^3}\epsilon^2 \nonumber\\
\delta  &=&3+\epsilon
  +\frac{(n^2+14n+60)}{2(n+8)^2}\epsilon^2 \nonumber\\
 \nu &=& \frac{1}{2} + \frac{n+2}{4(n+8)}\epsilon +
 \frac{n+2}{8(n+8)^3}(n^2+23n+60)\epsilon^2 \nonumber\\
\eta &=& \frac{n+2}{2(n+8)^2}\epsilon +
 \frac{n+2}{8(n+8)^4}(56n +272-n^2)\epsilon^2 \nonumber\\
\phi &=& 1 + \frac{n}{2(n+8)}\epsilon +
 \frac{n^2+24n+68}{4(n+8)^3}\epsilon^2
\end{eqnarray}
Here we list explicitly the values of the critical exponents
in the table \ref{table:one}.
It is easy to check the scaling law is approximately satisfied.

%\subsection{Observable prediction}

{\it Critical temperatures:} As already discussed in Ref. \cite{science},
the behavior of the SC transition
temperature $T_c$ and the AF transition temperature $T_N$
close to the bicritical point are governed by the $SO(5)$
bicritical exponent $\phi$.
In the neighborhood of the bi-critical
point, divergent quantities generally behave like:
\begin{eqnarray}
  \chi(T,g) \sim t^{-\gamma_5} X(g/t^\phi)
\end{eqnarray}
where $t=T-T_c(g=0)$ and $X(z)$ is a
scaling function, normalized such that $X(0)=1$. However, unlike
the usual scaling functions, $X(z)$ diverges at two points $z_2>0$
and $z_3<0$:
\begin{eqnarray}
  X(z) \sim (z-z_2)^{-\gamma_2} \ \ ; \ \ X(z) \sim (z-z_3)^{-\gamma_3}
\end{eqnarray}
Therefore, sufficiently close to the $SO(5)$ bicritical point,
$g/t^\phi\ll 1$, and the critical behavior is given by the
new $SO(5)$ exponent $\gamma_5$. Away from the $SO(5)$ bicritical point,
the divergence of physical quantities are determined by the
divergence of $X(z)$. For $g>0$, the critical temperature is
given by $g/t^\phi=z_2$, or
\begin{eqnarray}
T_c(g)=T_c(0) + A g^{1/\phi}
\end{eqnarray}
where $A$ is a constant. Similar arguments applies for the case of $g<0$.
This way, by measuring the precise values of both $T_c$ and $T_N$
close to the bicritical point, one can determine the value
of the crossover exponent $\phi$ and compare it with the
$SO(5)$ prediction of $\phi=1.314$.

{\it London penetration length:} Near the superconducting to normal
phase transition, this quantity scales like\cite{penetration}
\begin{eqnarray}
  \lambda \sim \rho_s^{-\frac{1}{2}}\sim \xi^{-(2-D)/2} \sim t^{-\nu/2}
\end{eqnarray}
This is a very interesting quantity, since
its critical behavior has already been measured by Kamal et al.
\cite{bonn} for YBCO
superconductors, and the exponent was found to be consistent with the
XY value of $\nu_2=0.655$. Here we suggest to measure the critical
behavior of $\lambda$ for doping levels ranging from optimal
to deeply underdoped regime. The $SO(5)$ theory predicts that
data for all doping levels $x$ can be fit into a single scaling
curve:
\begin{eqnarray}
  \chi(T,x) \sim t^{-\nu_5/2} Y(x/t^\phi)
\end{eqnarray}
where $Y(0)=1$ and it diverges near $z_2>0$ as:
\begin{eqnarray}
  Y(z) \sim (z-z_2)^{-\nu_2/2}
\end{eqnarray}
If one can get sufficiently close to the bicritical regime,
one can determine both $\nu_5$ and $\phi$ and compare with
the $SO(5)$ predictions of $\nu_5=0.714$ and $\phi=1.314$.
Together with these precisely predicted values, the
fitting into a single scaling curve for all doping levels
provides a highly nontrivial quantitative test of the
$SO(5)$ theory.

Other static quantities should all follow similar scaling
relations close to the bicritical point.

%\section{Dynamic exponent}
%\subsection{Theory}
{\bf Dynamic exponent:}
Under the dynamic scaling hypothesis, the typical frequency
or the relaxation rate $\omega$ scales as
$ \omega \sim \xi^{-z}$.
Standard arguments in dynamical critical phenomena gives
$z=d/2$ generally, and $z=\phi/\nu$ near a bicritical
point\cite{dynamic}.

{\it Nuclear magnetic relaxation rate:} $1/T_1$ is given by the
following response function:
\begin{eqnarray}
1/T_1 = lim_{\omega\rightarrow 0} \frac{1}{\omega}
\int \frac{d^dk}{(2\pi)^d} \chi(k,\omega)
\end{eqnarray}
where $\chi(k,\omega)$ is the {\it imaginary part} of the
spin response function, which
near a critical point behaves like
\begin{eqnarray}
\chi(k,\omega) = \xi^{2-\eta} Y(\bar k, \bar \omega)
\end{eqnarray}
Here $Y(\bar k, \bar \omega)$ is a scaled, dimensionless
spin correlation function of the dimensionless variable
$\bar k = k\xi$ and $\bar \omega =\omega\xi^z$. Expressed
in terms of the rescaled variables,
\begin{eqnarray}
1/T_1 = \xi^{z-d+2-\eta} lim_{\bar\omega\rightarrow 0}
\frac{1}{\bar\omega}
\int \frac{d^d\bar k}{(2\pi)^d} Y(\bar k,\bar \omega)
\end{eqnarray}
from which we can see that the scaling behavior of
$1/T_1$ is given by:
\begin{eqnarray}
1/T_1 = \xi^{z-d+2-\eta} = t^{-x}
\end{eqnarray}
where $x=\nu(z-1-\eta)$. Applying the results of the
$\epsilon$ expansion listed in the previous table,
we obtain $x = \nu(z-1-\eta) =0.573 $
close to a $SO(5)$ bicritical point.
For a regular antiferromagnetic transition
of the $O(3)$ symmetry class, we obtain
$x=\nu(z-1-\eta) = 0.67(1.5-1-0.039)
=0.312.$

{\it Frequency-dependent conductivity:}
In the superconducting state $T<T_c$, and for low frequency, the
complex conductivity takes the form
\begin{eqnarray}
\sigma(\omega) \sim \frac{\rho_s}{-i\omega}
\end{eqnarray}
In the critical region, as we pointed out before,
$\rho_s \sim \xi^{2-d}$,
therefore, the dynamic conductivity scales as
\begin{eqnarray}
\sigma(\omega) \sim \xi^{2-d}/\omega \sim \omega^{-\frac{z+2-d}{z}}
\end{eqnarray}
at $T=T_c$ for low frequency.
At a $SO(5)$ biciritical point, the exponent is given by
$\frac{z+2-d}{z} =0.46$ compared with
$\frac{z+2-d}{z} =0.33$ for a ordinary superconductor to
normal transition in the $XY$ universality class.

%{\it DC conductivity:} According to above reasoning,
%the dc conductivity for $T\rightarrow T_c^+$ scales as
%\begin{eqnarray}
%\sigma(\omega=0) \sim t^{-\nu(z+2-d)}
%\end{eqnarray}
%At the $SO(5)$ biciritical point, $\nu(z+2-d) = 0.6$
%compared to $ 0.33 $ for the 3-D $XY$ model.

{\bf Experimental status:} Possibly due to chemical complications,
it is hard to reach a uniform state in the deeply underdoped
regime of the high $T_c$ superconductors. For this reason, the
existence of a bicritical in the high $T_c$ superconductors has
neither been discovered nor refuted. One of the greatest
experimental challenge in this field is to prepare better and
more uniform materials in the deeply underdoped regime. Given such
materials, it is most feasible to measure the critical properties of the
London penetration length by microwave cavity experiment. As clearly
demonstrated in this work, such system can provide definite
and quantitative test of the $SO(5)$ theory of high $T_c$
superconductivity.

Encouraging experimental evidence for a $SO(5)$ bicritical point
does exist in a class of 2D organic superconductors called $bedt$
salt. These material share most common physical properties
with the cuprates, and a AF to SC transition can be induced by
pressure. A bicritical point exists where $T_c$ and $T_N$
intersect each other. Kanoda and coworkers\cite{kanoda} measured the
$1/T_1$ rate both in the AF region and the bicritical region.
Murakami and Nagaosa\cite{organic} analyzed the experimental data.
The $1/T_1$ exponent in the AF region was measured to be
$x_{AF}=0.30$, compared with the theoretical prediction of $x_3=0.312$.
In the bicritical region, the experimental fit gives $x_{bi}=0.56$,
compared with the $SO(5)$ theoretical prediction of $x_5=0.573$.
This is the first experiment which directly measures the dimension
of the symmetry group close to a AF/SC bicritical point, and determines
$n$ to be close to $5$.

{\bf Conclusions:} $SO(5)$ theory makes precise and quantitative
predictions on the critical exponents near a bicritical point.
We strongly encourage experiments to be carried out in deeply
underdoped regime of high $T_c$ superconductors and to look for
the bicritical point. Measurement of the critical exponents associated
with various physical quantities can uniquely test the fundamental
validity of the $SO(5)$ theory, and can measure the number $5$ of the
the $SO(5)$ theory in a direct and unambiguous  fashion.

{\bf Acknowledgement:} We would like to thank Dr. S. Murakami and
N. Nagaosa for stimulating communications, their work directly
inspired ours. We would like to thank the organizers of the m2s
meeting for the invitation to contribute to this proceeding.
This work is supported by
the NSF under grant numbers DMR-9814289 and DMR-9400372. JP. Hu.
is supported by the Stanford Graduate Fellowship Program.

\hspace{1cm}

\begin{table}
\begin{tabular}{|c|c|c|c|c|}
n & 2 & 3 & 4 & 5 \\ \hline
$\alpha$ & -0.02 &-0.1 & -0.167 &-0.222 \\ \hline
$\beta$ & 0.36 & 0.377 & 0.391 & 0.402 \\ \hline
$\gamma$ & 1.3 & 1.347 & 1.385 & 1.418 \\ \hline
$\delta$ & 4.46 & 4.559 & 4.458 &4.458 \\ \hline
$\nu$ & 0.655 & 0.678 & 0.698 & 0.714 \\ \hline
$\eta$ & 0.039 & 0.039 & 0.038 & 0.037 \\ \hline
$\phi $& 1.16 & 1.22 & 1.27 & 1.314 \\ \hline
$z$  & 1.771 & 1.799 & 1.819 &1.840 \\
\end{tabular}
%\vspace{5pt}
\caption{Static and dynamic exponents for $n$ vector model}
\label{table:one}
\end{table}

\end{document}